\documentclass[twocolumn,showpacs,preprintnumbers,amsmath,amssymb]{revtex4}

\usepackage{graphicx}
\usepackage{dcolumn}
\usepackage{bm}
\usepackage{color}

\begin{document}

\preprint{}

\title{Zn and Ni doping effects on the low-energy spin excitations in La$_{1.85}$Sr$_{0.15}$CuO$_{4}$
}

\author{M. Kofu}
\affiliation{
Institute for Solid State Physics, University of Tokyo, Kashiwa 277-8581, Japan
}%
\affiliation{%
Department of Physics, Tohoku University, Sendai 980-8578, Japan
}%

\author{H. Kimura}
\affiliation{
Institute of Multidisciplinary Research for Advanced Materials, Tohoku University, Sendai 980-8577, Japan
}%

\author{K. Hirota}%
\affiliation{
Institute for Solid State Physics, University of Tokyo, Kashiwa 277-8581, Japan
}%

\date{\today}

\begin{abstract}
Impurity effects of Zn and Ni on the low-energy spin excitations 
were systematically studied in optimally doped La$_{1.85}$Sr$_{0.15}$Cu$_{1-y}$A$_{y}$O$_{4}$ (A=Zn, Ni) by neutron scattering.
Impurity-free La$_{1.85}$Sr$_{0.15}$CuO$_{4}$ shows a ``spin gap'' of $\omega =4$~meV
below $T_{\rm c}$ in the antiferromagnetic(AF) incommensurate spin excitation.
In Zn:$y=0.004$, the spin excitation shows a spin gap of 3~meV below $T_{\rm c}$.
In Zn:$y=0.008$ and Zn:$y=0.011$, however, the dynamical susceptibility $\chi ''$ at $\omega =3$~meV decrease below $T_{\rm c}$
and increase again at lower temperature, indicating an {\em in-gap} state.
In Zn:$y=0.017$, the low-energy spin state remains unchanged with decreasing temperature,
and elastic magnetic peaks appear below $\sim 20$~K then exponentially grow.
As for Ni:$y=0.009$ and Ni:$y=0.018$, the low-energy excitations below $\omega =3$~meV and 2~meV disappear below $T_{\rm c}$.
The temperature dependence of $\chi ''$ at $\omega =3$~meV, however, shows no upturn in contrast with Zn:$y=0.008$ and Zn:$y=0.011$,
indicating the absence of in-gap state.
In Ni:$y=0.029$, the magnetic signals were observed also at $\omega =0$~meV.
Thus the spin gap closes with increasing Ni.
Furthermore, as $\omega $ increases, the magnetic peak width broadens and 
the peak position shifts toward the magnetic zone center ($\pi$ $\pi$).
We interpret the impurity effects as follows:
Zn locally makes a non-superconducting {\em island} exhibiting the in-gap state in the superconducting {\em sea} with the spin gap.
Zn reduces the superconducting volume fraction, thus suppressing $T_{\rm c}$.
On the other hand, Ni primarily affects the superconducting sea, 
and the spin excitations become more dispersive and broaden with increasing energy,
which is recognized as a consequence of the reduction of energy scale of spin excitations.
We believe that the reduction of energy scale is relevant to the suppression of $T_{\rm c}$.

\end{abstract}

\pacs{Valid PACS appear here}
\maketitle

\section{\label{sec:intro}Introduction}

In past years, a large number of neutron scattering studies have suggested
that antiferromagnetic (AF) spin fluctuations are relevant to the mechanism of high-$T_{\rm c}$ superconductivity.
Significant features of AF spin correlations in exhaustively studied LSCO and YBCO are summarized as follows; 
(i) A well-defined gap with a gap energy $E_{\rm sg}$ opens below the superconducting transition temperature $T_{\rm c}$ 
in the energy spectrum, which is often called ``spin gap.''
Note that the spin gap in neutron scattering appears at temperature lower than that observed in NMR,
because neutron scattering detects not local spin behaviors but spatial structures of spins with a coherence.
As for LSCO, the spin gap is observed only around the optimum doping concentration $0.15 \leq x \leq 0.18$.~\cite{Yamada1995, Lee2000, Lee2003}
On the other hand, the spin gap of YBCO appears in the superconducting state and there exists a relationship $E_{\rm sg} \propto T_{\rm c}$
in the region $0.5 \leq x  \leq 0.95$.~\cite{Dai2001}
(ii) In both LSCO and YBCO, the spin excitations appear at incommensurate positions $Q=$~($\frac{1}{2}\pm \delta$ $\frac{1}{2}$ 0),
($\frac{1}{2}$ $\frac{1}{2}\pm \delta$ 0) in the high-temperature tetragonal (HTT) notation.
Additionally, the incommensurability $\delta $ is proportional to $T_{\rm c}$ in LSCO($x \leq 0.15$)~\cite{Yamada1998} 
and YBCO($x \leq 0.6$).~\cite{Dai2001}
However, it is ambiguous whether the incommensurate peaks of LSCO and that of YBCO have the same origin.
For example, the incommensurate peaks of YBCO are not clearly separated due to the broadened peak.
(iii) The magnetic excitation of YBCO is enhanced below $T_{\rm c}$ at a particular energy $E_r$($\propto T_{\rm c}$) 
and a position $Q=$~($\frac{1}{2}$ $\frac{1}{2}$ $l$), which is called a magnetic resonance peak.
The existence of the magnetic resonance peak in LSCO is still an open question.~\cite{Tranquada2004_a, Christensen2004}


%
To study whether a certain phenomenon is related to the superconductivity, reducing $T_{\rm c}$ is very effective.
There exist many ways for $T_{\rm c}$ reduction; for example, crystal structure, magnetic field, and impurity substitution.
Impurity substitution for Cu sites can be chosen so as to affect a particular property;
ions with a different valence change carrier doping rate in the CuO$_2$ plane, the ionic radius mismatch induces local distortions, 
and the spins of impurities affect magnetic correlations.
%
%
Thus, we can control a particular parameter to investigate which parameter is most related to $T_{\rm c}$.
Substitutions by divalent transition metal ions for Cu sites conserve the carrier doping level
and hardly influence on the macroscopic crystal structure,
which is reflected in the changes of lattice constants and structural phase transition temperature.
In particular, non-magnetic ion Zn$^{2+}$ (3d$^{10}$, $S=0$) and magnetic ion Ni$^{2+}$ (3d$^{8}$, $S=1$) are suitable to
understanding how magnetic interaction contributes to the high-$T_{\rm c}$ pairing mechanism.
%
It is known that the $T_{\rm c}$-suppression effect of Zn is stronger than that of Ni, unlike those for BCS superconductors.
Furthermore, magnetic susceptibility measurements show 
that local magnetic moments induced by Zn are larger than that by Ni.~\cite{Xaio1990}
NMR studies for YBCO indicate that there exist staggered moments on Cu sites around Zn,~\cite{Mahajan1994, Julien2000}
suggesting that Zn induces local magnetic moments around itself.
In scanning tunneling microscopy (STM) studies for impurity-doped Bi$_{2}$Sr$_{2}$CaCu$_{2}$O$_{8+\delta }$ (Bi2212)
near the impurity site, Zn induces an intense quasiparticle scattering resonance,~\cite{Pan2000}
while Ni makes quasiparticle state near the superconducting gap.~\cite{Hudson2001}
These results indicate that effects of Zn on the superconductivity and the magnetism are different from those of Ni.
%
For the study of impurity doping, we should choose a high-$T_{\rm c}$ superconductor 
which overall behaviors have already been revealed and has the single CuO$_2$ layer
in order to avoid the ambiguity for which Cu site impurities are substituted.
For this reason, LSCO is most suitable.


%

Systematic neutron scattering studies on impurity substitution had not been performed,
because neutron scattering studies need large and homogeneous single crystals.
It is even more difficult to prepare a single crystal with a correctly controlled impurity concentration.
Recently, however, Kimura {\it et al.} have systematically studied Zn doping effects 
on the low-energy AF spin excitations for optimally doped La$_{1.85}$Sr$_{0.15}$CuO$_{4}$ 
with various Zn concentrations, through strenuous efforts for crystal growth.~\cite{Kimura2003_a, Kimura2003_b}
%
%
They have revealed that Zn induces a novel AF spin state in the spin gap, which they call an {\em in-gap} state.
They also found that the induced state becomes more dominant and more static with increasing Zn.
Combined with STM and $\mu $SR studies which suggest that the superconductivity is excluded around Zn,~\cite{Pan2000, Nachumi1996}
they have concluded that the induced state appears in the non-superconducting region
and that the induced state regions are spatially separated from the superconducting regions which exhibit the spin gap state.
%

In this paper, we report Zn and Ni doping effects on the low-energy AF spin excitations and the superconducting transition.
Our purpose is to clarify the difference between the effects of Zn and those of Ni.
We performed neutron scattering and magnetic susceptibility measurements in La$_{1.85}$Sr$_{0.15}$Cu$_{1-y}$Zn$_{y}$O$_{4}$ (Zn:$y=0.004$, 0.008, 0.011, 0.017)
and La$_{1.85}$Sr$_{0.15}$Cu$_{1-y}$Ni$_{y}$O$_{4}$ (Ni:$y=0.009$, 0.018, 0.029).


\section{\label{sec:sample}sample preparation and experimental details}

Neutron scattering study for spin excitations requires large and spatially homogeneous single crystals due to weak signals.
Moreover, for the study of the impurity effect, 
it is essential to control impurity concentrations correctly without changing other properties,
for example, Sr concentration homogeneity and mosaicness.
We controlled the size, shape and growth direction of crystals for all samples
in order to unify experimental conditions and realize the quantitative comparisons of results among different samples.
%

Single crystals of La$_{1.85}$Sr$_{0.15}$CuO$_{4}$, La$_{1.85}$Sr$_{0.15}$Cu$_{1-y}$Zn$_{y}$O$_{4}$ (Zn:$y=0.004$, 0.008, 0.011, 0.017) and 
La$_{1.85}$Sr$_{0.15}$Cu$_{1-y}$Ni$_{y}$O$_{4}$ (Ni:$y=0.009$, 0.018, 0.029)
were grown using the traveling-solvent-floating-zone (TSFZ) method.
Feed rods were prepared by the ordinary solid-state reaction method.
Dried powders of La$_2$O$_3$, SrCO$_3$, CuO, ZnO, and NiO were weighed and mixed.
The mixed powder was ground and calcined at 900~$^{\rm o}$C for 20~h for four times in air.
After several grindings and calcinations of powder, we added extra CuO of 2~mol\% into the powder in order to compensate the loss of Cu during the crystal growth.
The powder was formed into a cylindrical rod and the rod was hydrostatically pressed.
We sintered the rod at 1250$^{\rm o}$C for 24~h in air.
The solvent material with a composition of La:Sr:Cu:Zn(Ni)=1.85:0.15:(4-$y$):$y$ for each $y$ was prepared.
We use 0.3~g of the solvent and a La$_{1.85}$Sr$_{0.15}$Cu$_{1-y}$A$_{y}$O$_{4}$ (A=Zn, Ni) single crystals as a seed rod.
We controlled the growth direction of crystals by using the seed rod which the direction of cylinder is $a$-axis.
The TSFZ operation was carried out using infrared image furnaces (NEC Machinery Co., SC-4 and SC-35HD) with small two halogen lamps and double ellipsoidal mirrors.
To cut off reflections of the light from high angles, a silica tube was partially covered with aluminum foils.
This yields a sharp temperature gradient in the direction of crystal growth, which helps to stabilize the melting zone.
The growth rate was kept constant at 1.0~mm/h, and 
both the feed and seed rods were rotated 30 and 15~rpm, respectively.
We flowed mixed gas of argon and oxygen with the flow rate of 100~cm$^3$/min.(Ar:20~cm$^3$/min., O$_2$:80~cm$^3$/min.).
Crystal growth conditions for all the samples were kept same and we maintained the conditions for 100~hours during crystal growth.
After crystal growth, we put the crystal rod in water for 1~week, to check the inclusion of La$_2$O$_3$ which reacts with water and changes to La(OH)$_3$.
The appearance of La(OH)$_3$ break the crystal, because the size of La(OH)$_3$ is larger than that of La$_2$O$_3$.
Finally, to eliminate oxygen deficiencies, all the crystals were annealed under oxygen gas flow at 900~$^{\rm o}$C for 50~hours, 
cooled to 500~$^{\rm o}$C at a rate of 10~$^{\rm o}$C/h and annealed at 500~$^{\rm o}$C for 50~hours.

The concentrations of La, Sr, Cu, Zn and Ni ions were precisely estimated by the inductively coupled plasma (ICP) analysis
using Shimadzu ICPS-7500.
Within the detection limit of the ICP analysis, no observable impurities were detected.
We checked the spatial homogeneity of Sr, Zn and Ni ions by comparing the values 
for small portions of crystals taken at different points of each sample,
and we confirm the macroscopic homogeneity of composition.
The obtained values of Sr, Zn and Ni concentrations are summarized in Table~\ref{tab:characterization}.
The values of Sr concentration for all the sample are $x \sim 0.15$ within the measurement error
and those of impurity concentrations are substantially larger than the error.
These results show that we could control the impurity doping rate with keeping the Sr concentration.
%
We also determined the structural phase transition temperature $T_{\rm d1}$ by neutron diffraction measurement.
As temperature decreases, the crystal structure changes from high-temperature tetragonal (HTT) to low-temperature orthorhombic (LTO) at $T_{\rm d1}$.
The (072)$_{\rm ortho}$ superlattice peak in the LTO notation appears below $T_{\rm d1}$ and we measured temperature dependence of the (072)$_{\rm ortho}$ peak intensity.
As seen in Table~\ref{tab:characterization}, the obtained values of $T_{\rm d1}$ for all the samples are almost same 
and consistent with that of LSCO ($x=0.15$).~\cite{Yamada1995}
$T_{\rm d1}$ is very sensitive to the Sr concentration and the $T_{\rm d1}$ reduction rate is $\sim 22$K$/x$ at.\%.
Hence, these results show that the Sr concentration for all the samples are $x \sim 0.15$
and Sr ions are doped homogeneously into the crystals.
In the Zn-doped samples, $T_{\rm d1}$ becomes somewhat high with increasing the doping contents.
On the other hand, $T_{\rm d1}$ of the Ni-doped samples is almost equal to that of the impurity-free sample.
Gaojie {\em et al.} reported that the crystal structure of Zn-doped LSCO($x=0.15$, Zn:$y\geq 0.15$) is orthorhombic in the room temperature,
while that of Ni-doped LSCO($x=0.15$, Ni:$y\leq 0.30$) remain tetragonal.~\cite{Gaojie1997}
Their results are consistent with our results;
though Zn slightly increases $T_{\rm d1}$, a small amount of impurities hardly affects the averaged crystal structure.
%
Furthermore, we measured rocking curves of crystals by neutron diffraction.
The full width half maximum (FWHM) are $\sim 0.15^{\rm o}$ for all the samples, suggesting a good crystallinity.

\begin{table}[t]
\caption{\label{tab:characterization}
Sr, Zn and Ni concentrations estimated by ICP analysis
and $T_{\rm d1}$ determined by netron diffraction in La$_{1.85}$Sr$_{0.15}$Cu$_{1-y}$A$_{y}$O$_{4}$(A=Zn, Ni).
}
\begin{ruledtabular}
\begin{tabular}{cccc}
sample & Sr $x$ & Zn or Ni $y$  & $T_{\rm d1}$ \\
\hline
$y=0$ & 0.142(8) & - & 186~K \\ \hline
Zn:$y=0.004$ & 0.144(3) & 0.0044(4) &  \\
Zn:$y=0.008$ & 0.147(4) & 0.0083(3) & 192~K \\
Zn:$y=0.011$ & 0.144(7) & 0.0112(5) &  \\
Zn:$y=0.017$ & 0.145(2) & 0.0170(7) & 202~K \\ \hline
Ni:$y=0.009$ & 0.145(7) & 0.0090(3) &  \\
Ni:$y=0.018$ & 0.144(6) & 0.0185(5) & 184~K \\
Ni:$y=0.029$ & 0.142(6) & 0.0287(8) & 182~K \\
\end{tabular}
\end{ruledtabular}
\end{table}

Inelastic neutron scattering experiments were performed with the Tohoku University triple-axis spectrometer (TOPAN) installed at the JRR-3M reactor
in the Japan Atomic Energy Research institute (JAERI).
The final energy was fixed at $E_f=13.5$~meV using the (002) reflection of a pyrolytic graphite analyzer.
The horizontal collimator sequence were 40$'$-100$'$-S-60$'$-80$'$ (a low-resolution condition) and 40$'$-30$'$-S-30$'$-80$'$ (a high-resolution condition),
where S denotes the sample position.
A pyrolytic graphite filter and a sapphire filter were placed to reduce higher order reflections and fast neutrons.
To compare magnetic signals quantitatively, these experimental conditions were unified throughout the experiments.
Elastic neutron scattering experiments were performed using TOPAN for Zn-doped sample and 
ISSP triple axis spectrometer PONTA also in JRR-3M for Ni-doped sample.
We selected the incident and final energies of 13.5~meV with the horizontal collimator sequence of 40$'$-30$'$-S-30$'$-80$'$ for the Zn-doped samples,
and 14.7~meV with 40$'$-40$'$-S-40$'$-80$'$ for the Ni-doped samples.
In order to increase the sample volume, the columnar-shaped crystals (typical size of each crystal was 5mm$\phi \times 25$mm) were assembled
and mounted in the ($hk$0) zone.
Since we controlled the growth axis parallel to the $a$-axis, 
all the crystal arrangements were kept identical.
In this paper, we denote the crystallographic indices by the HTT notation if no specification is given.
The crystals were mounted in alminum containers in which He gas was charged to act as a heat exchanger.
$^4$He-closed cycle refrigerators were used to cool the samples down to 4~K or 10~K for inelastic scattering measurements,
and a top-loading liquid-$^4$He cryostat was used to cool down to 1.5~K for elastic measurements.

\section{\label{sec:results}results}

\subsection{\label{sec:results_1}$T_{\rm c}$ suppression by impurity-doping}

\begin{figure}
\begin{center}
\includegraphics[width=0.75\hsize]{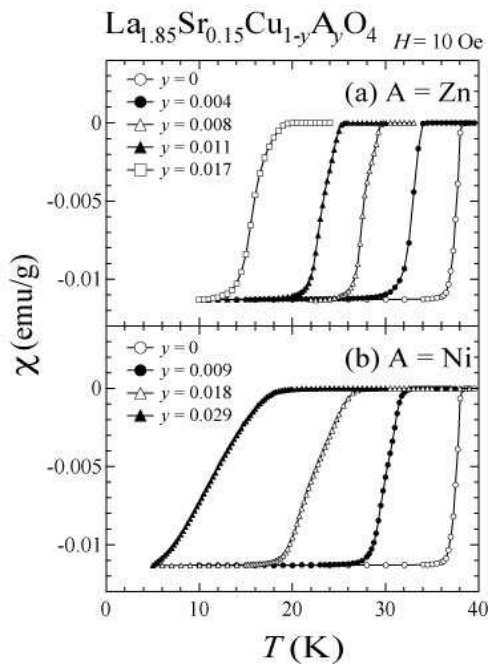}
\caption{
Shielding signals of (a)~La$_{1.85}$Sr$_{0.15}$Cu$_{1-y}$Zn$_{y}$O$_{4}$ (Zn:$y=0$, 0.004, 0.008, 0.011, 0.017) and 
(b)~La$_{1.85}$Sr$_{0.15}$Cu$_{1-y}$Ni$_{y}$O$_{4}$ (Ni:$y=0$, 0.009, 0.018, 0.029)
measured under a magnetic field of 10~Oe.
The values of susceptibility for all samples are scaled to $-\frac{1}{4\pi }$(emu).}
\label{fig:Tc}
\end{center}
\end{figure}

To estimate the $T_{\rm c}$ reduction by impurity-doping, 
we measured the magnetic susceptibilities under zero-field cooling using a SQUID magnetometer.
Figure~\ref{fig:Tc} shows the shielding signals of the impurity-doped LSCO samples,
and the signals are scaled so that the transition widths $\Delta T_{\rm c}$ are compared.
We define $T_{\rm c}$ as the temperature at which the shielding signal is half of the low-temperature saturated value,
and $\Delta T_{\rm c}$ is defined as a difference between two temperatures where the signal is 10\% and 90\% of the low-temperature value.
The values of $T_{\rm c}$ and $\Delta T_{\rm c}$ are listed in Table~\ref{tab:Tc}.
The values of $T_{\rm c}$ are in good agreement with those of previous studies,~\cite{Xaio1990}
and the reduction rates of $T_{\rm c}$ are 12.5~K/\% for Zn and 8.7~K/\% for Ni.
In addition, $\Delta T_{\rm c}$ is almost constant regardless of the Zn concentration,
while $\Delta T_{\rm c}$ broadens with increasing Ni.

\begin{figure}
\begin{center}
\includegraphics[width=0.55\hsize]{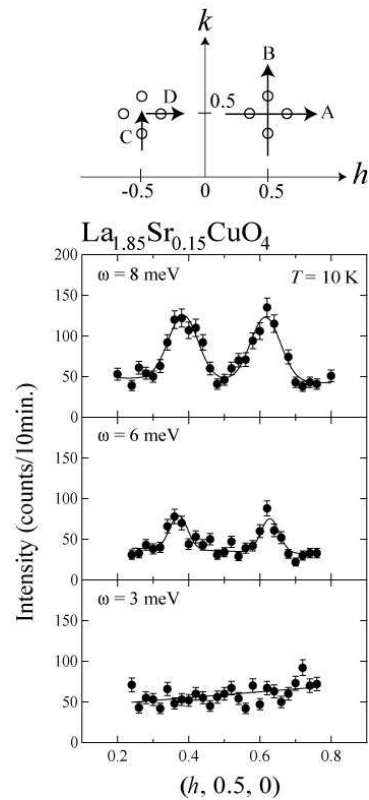}
\caption{
(Top) : Schematic drawing of scan trajectories. Open circles denote magnetic incommensurate peaks.
(Bottom) : Peak profiles along the $h$ direction (trajectory A) for La$_{1.85}$Sr$_{0.15}$CuO$_{4}$ at $\omega =3$, 6 and 8~meV.
Solid lines are the result of fits with assuming two equivalent Gaussian peaks at (0.5$\pm \delta $, 0.5, 0) and a linear background.
}
\label{fig:profiles(free)} 
\end{center}
\end{figure}

\begin{figure*}
\begin{center}
\includegraphics[width=\hsize]{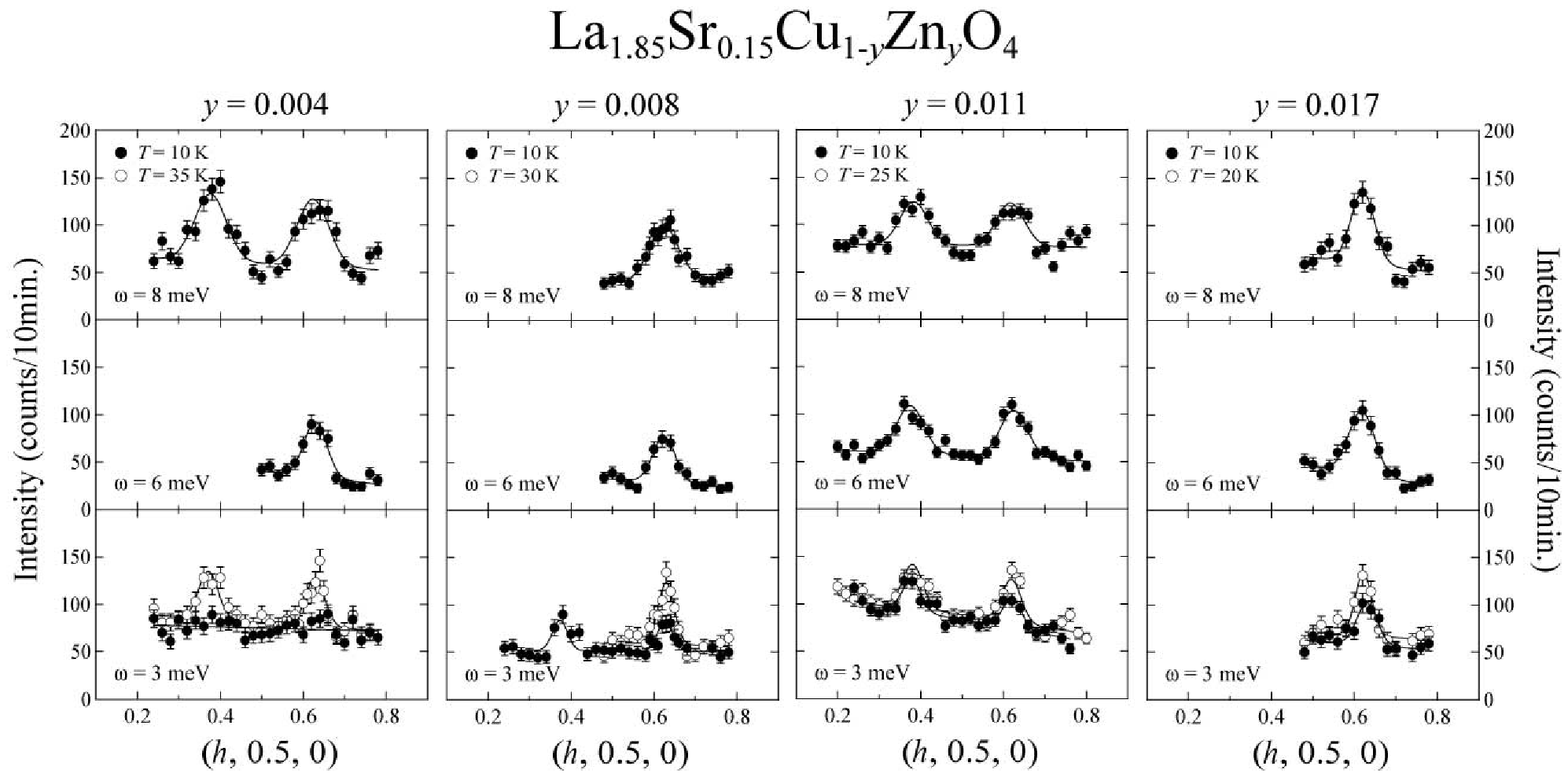}
\caption{
Peak profiles along the $h$ direction (trajectory A in Fig.~\ref{fig:profiles(free)}) 
for La$_{1.85}$Sr$_{0.15}$Cu$_{1-y}$Zn$_{y}$O$_{4}$ (Zn:$y=0.004$, 0.008, 0.011, 0.017) at $\omega =3$, 6 and 8~meV.
Filled and open circles correspond to the data below and above $T_{\rm c}$.
Solid lines are the fitting results by a Gaussian function with assuming two equivalent Gaussian peaks at (0.5$\pm \delta $, 0.5, 0) and a linear background.
We show the intensities divided by 2 for Zn:$y=0.011$, 
because the volume of sample with Zn:$y=0.011$ is approximately twice as large as those of the other Zn-doped samples.
}
\label{fig:Zn_profile_low} 
\end{center}
\end{figure*}

\begin{figure*}
\begin{center}
\includegraphics[width=0.784\hsize]{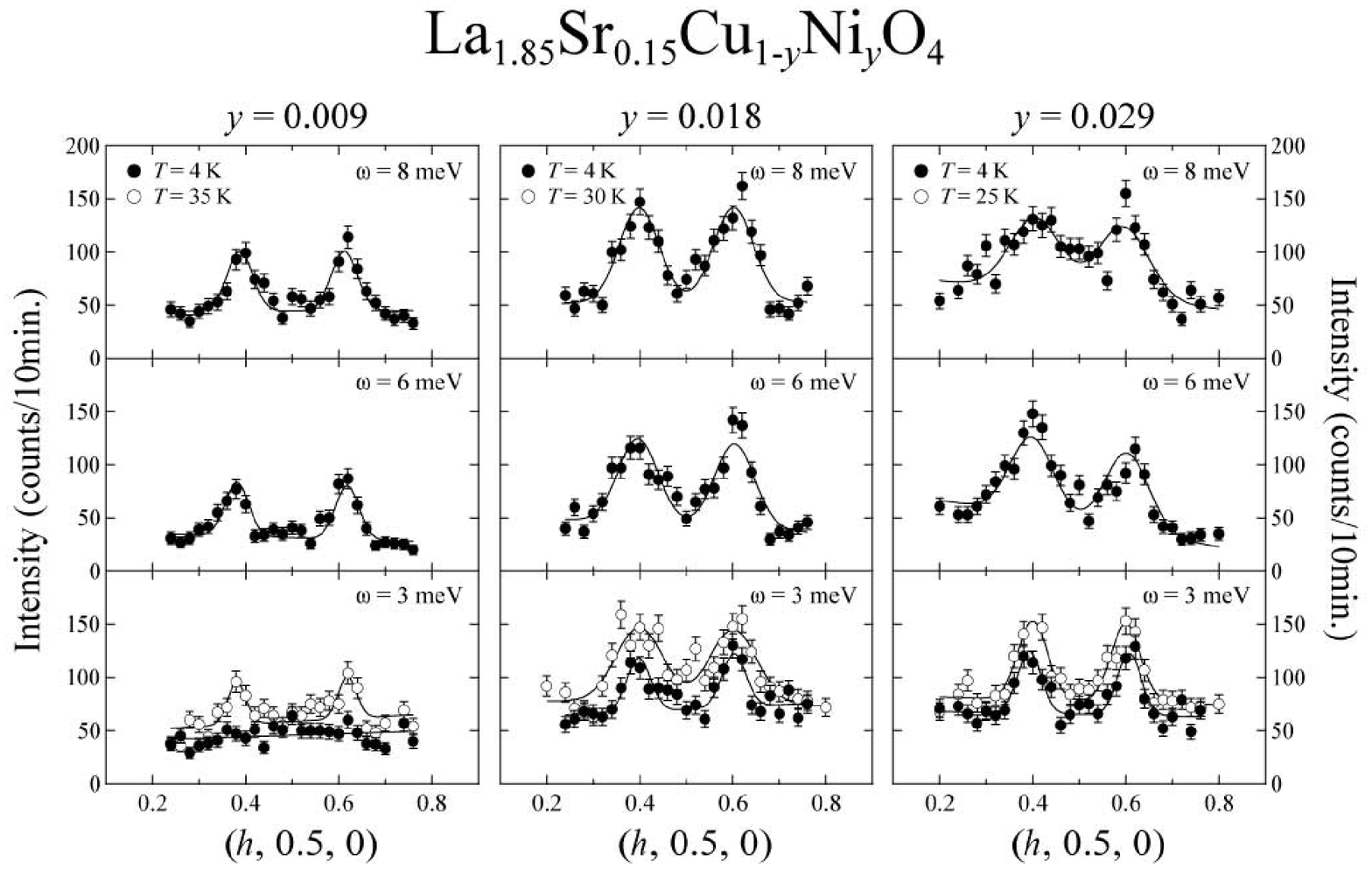}
\caption{
Peak profiles along the $h$ direction (trajectory A in Fig.~\ref{fig:profiles(free)}) 
for La$_{1.85}$Sr$_{0.15}$Cu$_{1-y}$Ni$_{y}$O$_{4}$ (Ni:$y=0.009$, 0.018, 0.029) at $\omega =3$, 6 and 8~meV.
Filled and open circles correspond to the data below and above $T_{\rm c}$.
Solid lines are the fitting results assuming two equivalent Gaussian peaks at (0.5$\pm \delta $, 0.5, 0) and a linear background.
}
\label{fig:Ni_profile_low} 
\end{center}
\end{figure*}

\begin{figure}
\begin{center}
\includegraphics[width=0.6\hsize]{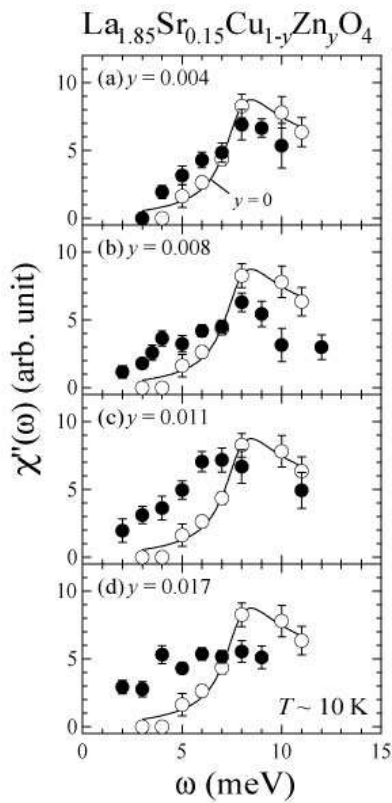}
\caption{
Energy dependence of the $Q$-integrated $\chi ''(\omega )$
for La$_{1.85}$Sr$_{0.15}$Cu$_{1-y}$Zn$_{y}$O$_{4}$ with (a)~Zn:$y=0.004$, (b)~Zn:$y=0.008$, (c)~Zn:$y=0.011$ and (d)~Zn:$y=0.017$
at $T \sim 10$~K.
Open circles are the data for the impurity-free sample and solid lines are guides to the eye.
}
\label{fig:kai(w)_Zn}
\end{center}
\end{figure}

\begin{figure}
\begin{center}
\includegraphics[width=0.6\hsize]{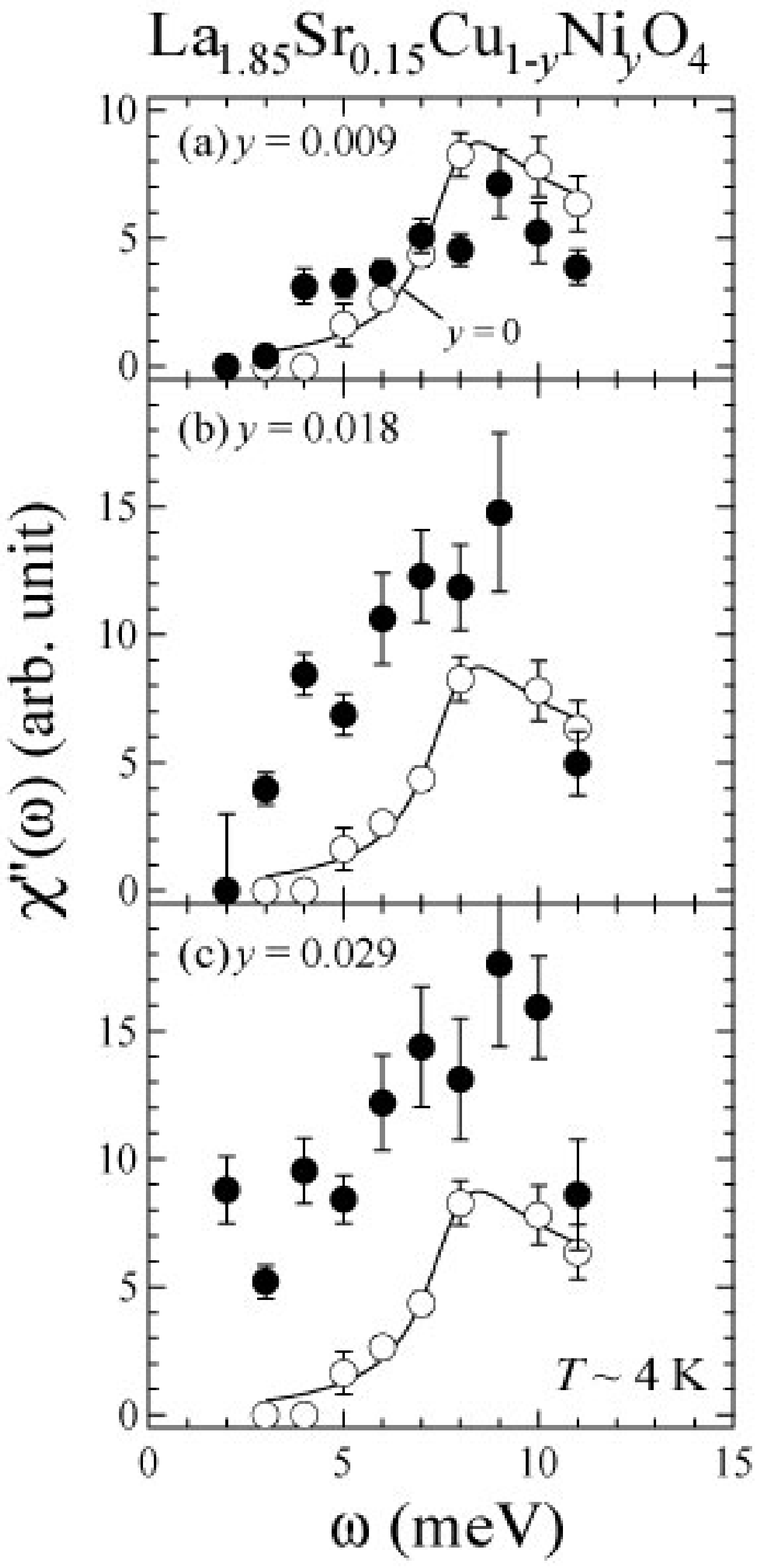}
\caption{
Energy dependence of the $Q$-integrated $\chi ''(\omega )$
for La$_{1.85}$Sr$_{0.15}$Cu$_{1-y}$Ni$_{y}$O$_{4}$ with (a)~Ni:$y=0.009$, (b)~Ni:$y=0.018$ and (c)~Ni:$y=0.029$
at $T \sim 4$~K.
Open circles are the data for the impurity-free sample and solid lines are guides to the eye.
}
\label{fig:kai(w)_Ni}
\end{center}
\end{figure}

\begin{table}[b]
\caption{\label{tab:Tc}
$T_{\rm c}$ and $\Delta T_{\rm c}$ estimated by shielding signals.
}
\begin{ruledtabular}
\begin{tabular}{ccc}
sample & $T_{\rm c}$(midpoint) & $\Delta T_{\rm c}$ \\
\hline
$y=0$ & 36.8~K & 1.7~K \\ \hline
Zn:$y=0.004$ & 32.6~K & 2.2~K \\
Zn:$y=0.008$ & 27.8~K & 2.5~K \\
Zn:$y=0.011$ & 23.2~K & 2.9~K \\
Zn:$y=0.017$ & 16.0~K & 3.8~K \\ \hline
Ni:$y=0.009$ & 30.0~K & 2.8~K \\
Ni:$y=0.018$ & 22.5~K & 5.9~K \\
Ni:$y=0.029$ & 11.6~K & 9.2~K \\
\end{tabular}
\end{ruledtabular}
\end{table}

One may consider that the broadening of $T_{\rm c}$ is caused by a spatial inhomogeneity of Ni concentration.
As for Ni:$y=0.029$, we performed two measurements
to investigate the inhomogeneity of Ni.
We cut the crystal into several portions with the size of 1~mm$^3$ and evaluated the Ni concentration of each portion by ICP analysis.
The obtained values are identical within the measurement error, and the error corresponds to $\Delta T_{\rm c}$ of 2~K.
This result indicates no macroscopic inhomogeneity of Ni in the crystal.
In addition, to check the microscopic inhomogeneity, we performed small angle neutron scattering measurements using the T1-2 double-axis spectrometer
attached to a thermal guide in JRR-3M, which is owned by IMR, Tohoku University.
If clusters of Ni ions are produced in a CuO$_2$ plane, signals should appear in the small angle region corresponding to a characteristic scale of the cluster size.
However, no such signal was detected in the range $0.1 \leq Q \leq  1$~\AA$^{-1}$.
If signals appear in the smaller angle region, one cluster contains Ni ions more than 30 and the mean distance between clusters exceeds 100\AA.
This situation seems improper,
because such a segregation should introduce two superconducting transition temperatures.
As shown in Fig.~\ref{fig:Tc}, though the superconducting transition broadens, the temperature dependence show no two-step behavior.
Besides, as listed in Table~\ref{tab:characterization}, the structural transition temperature $T_{\rm d1}$ does not change by Ni-doping.
If clusters of Ni ions exist, the averaged crystal structure should be affected.
Thus, we believe that there is no cluster of Ni in the crystal.
These behaviors, the broadening of $T_{\rm c}$ and no change of $T_{\rm d1}$, are already reported by Churei {\it et al.}~\cite{Churei2003}
and are consistent with our results.
%
We conclude that there is no microscopic nor macroscopic inhomogeneity of Ni ions
and that the broadening of $\Delta T_{\rm c}$ is an intrinsic effect of Ni-doping.


\subsection{\label{sec:results_2}Energy spectra of spin excitations}

Figure~\ref{fig:profiles(free)} shows constant-energy spectra of the spin excitations at $\omega =3$, 6, 8~meV for impurity-free La$_{1.85}$Sr$_{0.15}$CuO$_{4}$.
The scan through two incommensurate peaks corresponds to the trajectory~A illustrated at the top of Fig.~\ref{fig:profiles(free)}.
At 10~K, a clear spin gap was observed and is consistent with the previous studies.~\cite{Yamada1995, Lee2000}
In Figs.~\ref{fig:Zn_profile_low} and \ref{fig:Ni_profile_low}, we show $Q$ profiles at $\omega =3$, 6, 8~meV
for Zn:$y=0.004$, 0.008, 0.011, 0.017 and Ni:$y=0.009$, 0.018, 0.029,
measured below $T_{\rm c}$ (filled circles) and just above $T_{\rm c}$ (open circles).
The trajectories of scans and experimental conditions are identical to those for impurity-free LSCO.

In Zn:$y=0.004$ and Ni:$y=0.009$, the spin excitations at $\omega =3$~meV disappear at low temperatures, 
suggesting a spin gap opening analogous to impurity-free LSCO.
While, in Zn:$y \geq 0.008$ and Ni:$y \geq 0.018$, the magnetic signals at $\omega =3$~meV appear at low temperatures.
These results suggest that both Zn and Ni doping yield the appearance of low-energy spin excitations
and that the effect of Ni is weaker that that of Zn.
In Ni:$y=0.018$ and Ni:$y=0.029$, as $\omega $ increases, the magnetic peak intensity increases and the peak width apparently broadens.
The effect becomes marked with increasing Ni.
On the other hand, Zn does not affect the peak profiles of spin excitations.

Figures~\ref{fig:kai(w)_Zn} and \ref{fig:kai(w)_Ni} show energy spectra of the $Q$-integrated dynamical magnetic susceptibility $\chi ''(\omega )$ at low temperatures
for both the Zn and Ni doped samples.
The open circles and solid lines in the figures denote $\chi ''(\omega )$ for impurity-free LSCO.
Here $\chi ''(Q, \omega )$ corresponds to the dynamical structure factor $S(Q, \omega )$ corrected with the thermal population factor,
\begin{equation}
S(Q, \omega ) = \frac{1}{1-e^{-\hbar\omega /k_BT}}\chi ''(Q, \omega).
\end{equation}
Assuming that magnetic excitations consist of four equivalent incommensurate magnetic rods running along $l$,
we have estimated $\int S(Q, \omega ) dQ_{\delta}$, where $\int dQ_{\delta}$ denotes the integration of magnetic signals around one incommensurate magnetic peak.
Magnetic signals are fitted to Gaussian peaks and the fitting error corresponds to the error bar in Fig.~\ref{fig:Zn_profile_low}.
We also normalize the estimated $\chi ''(\omega )$ by the intensity of transverse acoustic phonons at (2, -0.19, 0)$_{\rm ortho}$
which should be proportional to the effective sample volume.

In the impurity-free sample, $\chi ''(\omega )$ has a maximum at $\sim 8$~meV (=$\omega _{\rm max}$), reflecting the gap structure.
In Zn:$y=0.004$ and Zn:$y=0.008$, the value of $\omega _{\rm max}$ remains unchanged,
but $\chi ''(\omega_{\rm max} )$ decrease and $\chi ''(\omega )$ around 3~meV increases instead.
These results suggest that the spin gap state decreases and an additional low-energy spin state appears.
In Zn:$y=0.011$, $\omega _{\rm max}$ shifts toward the lower energy.
At last, in Zn:$y=0.017$, the spin gap entirely vanishes,
because of the reduction of spin gap energy and the development of the additional state.



$\chi ''(\omega )$ of Ni:$y=0.009$ shows a behavior similar to that of Zn:$y=0.004$;
$\chi ''(\omega )$ above $\sim 7$~meV decreases while $\chi ''(\omega )$ below $\sim 7$~meV increases as compared to $y=0$.
In Ni:$y=0.018$, the low-energy spin excitations are observable at $\omega =3$~meV, but that below $\omega =2$~meV disappears.
The spin excitations above 2~meV are observed in Ni:$y=0.029$, showing a tendency of gap closing.
%
%
One of the most remarkable results is that the amplitude of $\chi ''(\omega )$ for Ni:$y=0.018$ and Ni:$y=0.029$ is larger than those for $y=0$ and Ni:$y=0.009$.

\begin{figure}
\begin{center}
\includegraphics[width=0.6\hsize]{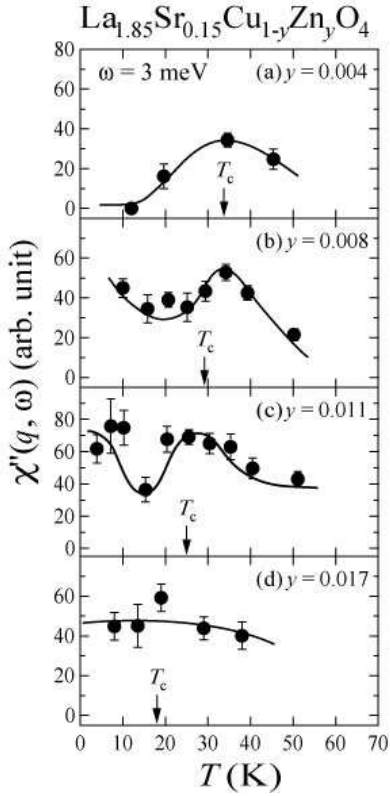}
\caption{
Temperature dependence of $\chi ''(Q, \omega)$
for La$_{1.85}$Sr$_{0.15}$Cu$_{1-y}$Zn$_{y}$O$_{4}$ with (a)~Zn:$y=0.004$, (b)~Zn:$y=0.008$, (c)~Zn:$y=0.011$ and (d)~Zn:$y=0.017$ at $\omega =3$~meV.
$\chi ''(Q, \omega)$ denote the fitted peak amplitude corrected with the thermal population factor.
Solid lines are guides to the eye and arrowheads indicate the onset values of $T_{\rm c}$.
}
\label{fig:kai(T-dep)_Zn}
\end{center}
\end{figure}

\begin{figure}
\begin{center}
\includegraphics[width=0.6\hsize]{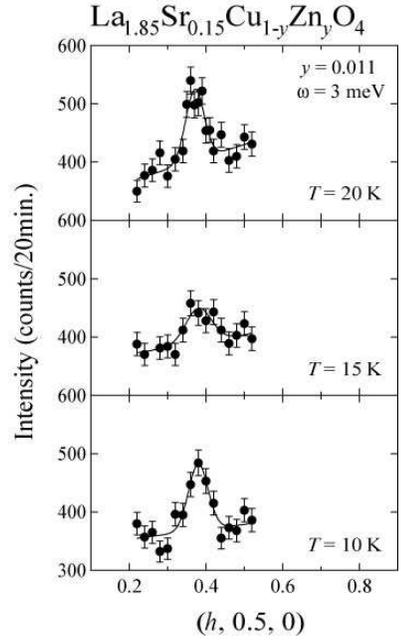}
\caption{
Peak profiles along the $h$ direction (trajectory A in Fig.~\ref{fig:profiles(free)}) 
for La$_{1.85}$Sr$_{0.15}$Cu$_{1-y}$Zn$_{y}$O$_{4}$ (Zn:$y=0.011$) at $T=10$, 15 and 20~K.
All the data were taken at $\omega =3$~meV.
Solid lines are the fitting results assuming a Gaussian peak and a linear backgound.
}
\label{fig:profile_Zn11}
\end{center}
\end{figure}

\begin{figure}
\begin{center}
\includegraphics[width=0.6\hsize]{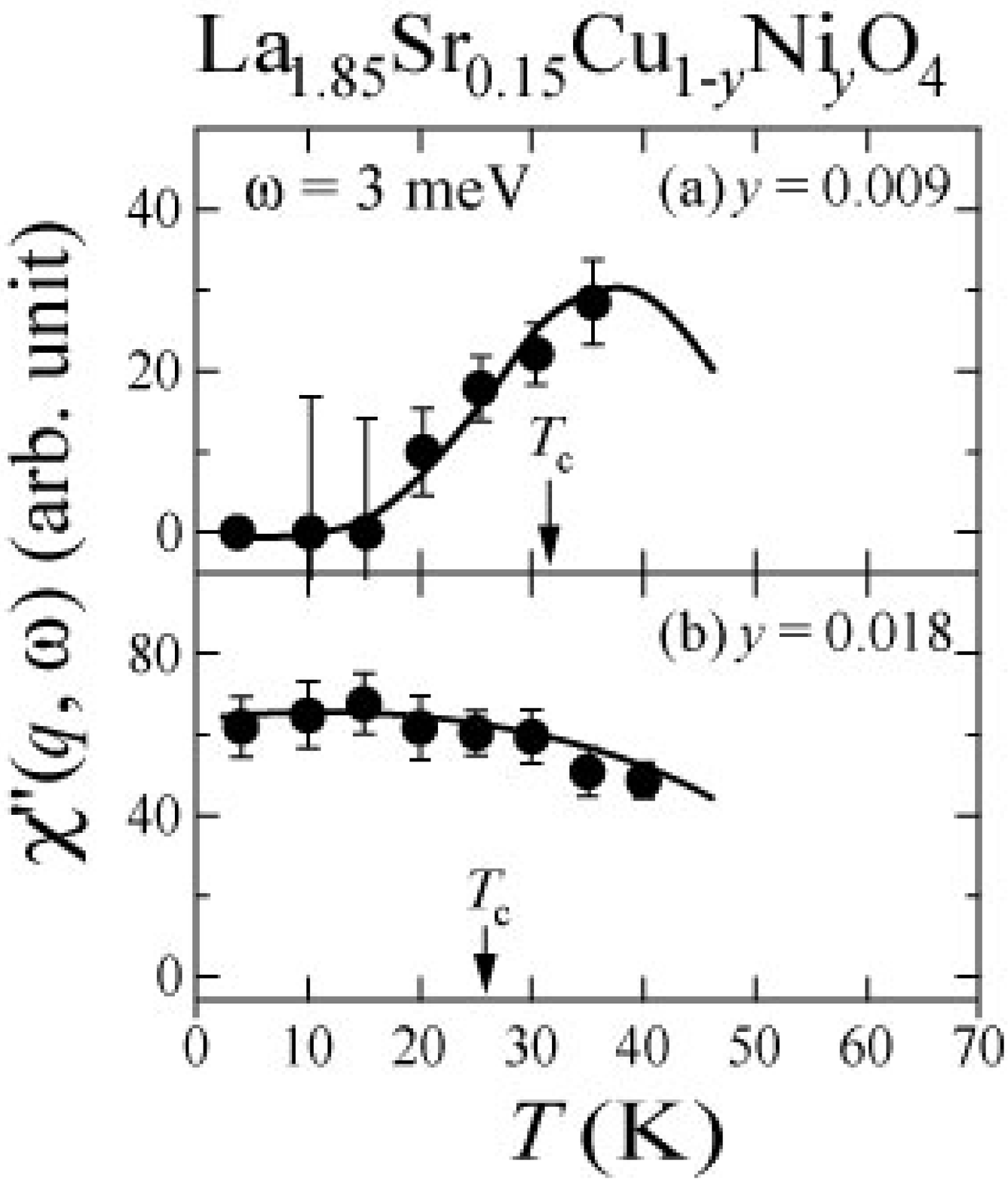}
\caption{
Temperature dependence of $\chi ''(Q, \omega)$
for La$_{1.85}$Sr$_{0.15}$Cu$_{1-y}$Ni$_{y}$O$_{4}$ with (a)~Ni:$y=0.009$ and (b)~Ni:$y=0.018$ at $\omega =3$~meV.
$\chi ''$($Q, \omega$) denotes the fitted peak amplitude corrected with the thermal population factor.
Solid lines are guides to the eye and arrowheads indicate the onset values of $T_{\rm c}$.
}
\label{fig:kai(T-dep)_Ni}
\end{center}
\end{figure}

\begin{figure}[t]
\begin{center}
\includegraphics[width=0.7\hsize]{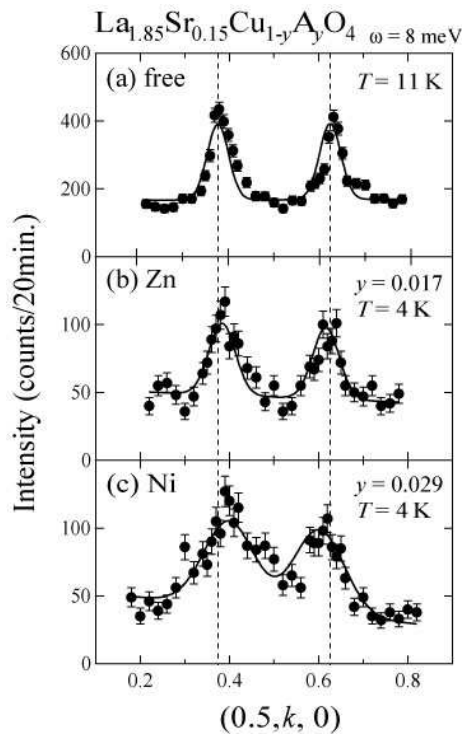}
\caption{
$Q$ profiles along the $k$ direction (trajectory B in Fig.~\ref{fig:profiles(free)}) 
for (a)~La$_{1.85}$Sr$_{0.15}$CuO$_{4}$, (b)~La$_{1.85}$Sr$_{0.15}$Cu$_{1-y}$Zn$_{y}$O$_{4}$ (Zn:$y=0.017$)
and (c)~La$_{1.85}$Sr$_{0.15}$Cu$_{1-y}$Ni$_{y}$O$_{4}$ (Ni:$y=0.029$) 
at $\omega =8$~meV under the high resolution condition.
These data were taken at low temperature.
Solid lines are the result of fits with assuming two equivalent Gaussian peaks at (0.5, 0.5$\pm \delta $, 0).
Vertical dashed lines denote peak positions for (a)~La$_{1.85}$Sr$_{0.15}$CuO$_{4}$  estimated by fitting.
}
\label{fig:profile(8meV)_high}
\end{center}
\end{figure}

\begin{figure}
\begin{center}
\includegraphics[width=0.7\hsize]{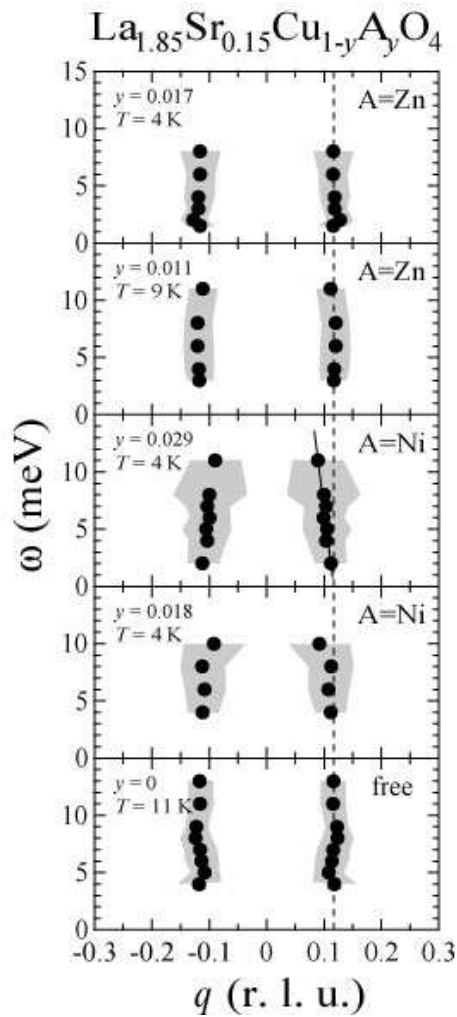}
\caption{
Spin excitations 
for $y=0$, La$_{1.85}$Sr$_{0.15}$Cu$_{1-y}$Zn$_{y}$O$_{4}$ (Zn:$y=0.011$, 0.017)
and La$_{1.85}$Sr$_{0.15}$Cu$_{1-y}$Ni$_{y}$O$_{4}$ (Ni:$y=0.018$, 0.029) 
at low temperature.
$q$ denotes the propagation vector of spin correlation; $q=0$ means the magnetic zone center $Q=$(0.5, 0.5, 0).
Closed circles and shaded regions represent the peak position and FWHM of incommensurate magnetic signals estimated from curve-fitting.
Dashed lines show a mean value of peak positions for the impurity-free sample.
A solid line indicates a guide to the eye of dispersion for Ni:$y=0.029$.
}
\label{fig:dispersion}
\end{center}
\end{figure}

\begin{figure}
\begin{center}
\includegraphics[width=0.75\hsize]{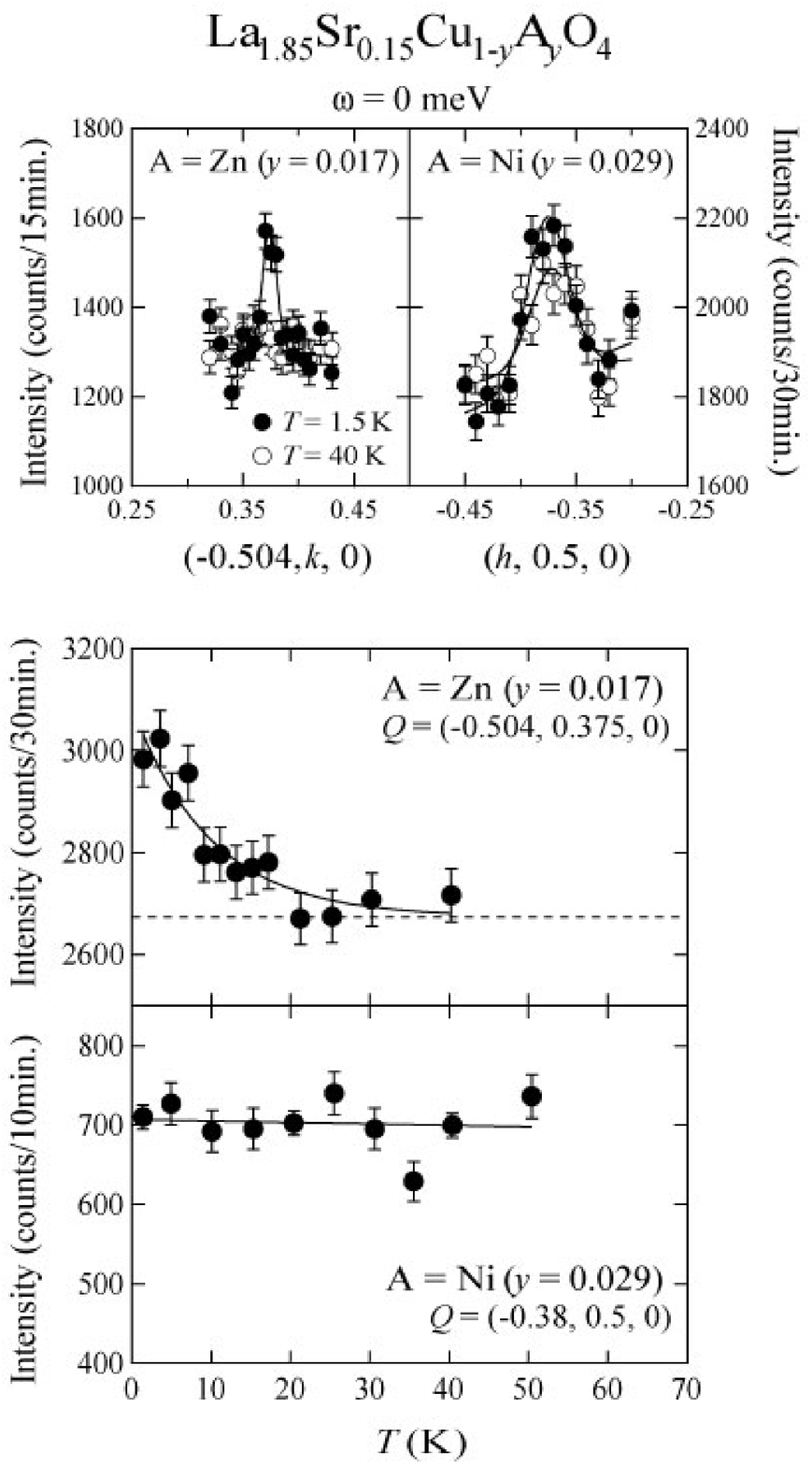}
\caption{
The incommensurate elastic peak intensities as a function of temperature
for La$_{1.85}$Sr$_{0.15}$Cu$_{1-y}$Zn$_{y}$O$_{4}$ (Zn:$y=0.017$)
and La$_{1.85}$Sr$_{0.15}$Cu$_{1-y}$Ni$_{y}$O$_{4}$ (Ni:$y=0.029$).
The top figures show $q$ profiles along the $k$ direction (trajectory C in Fig.~\ref{fig:profiles(free)}) for the Zn-doped sample and 
the $h$ direction (trajectory D in Fig.~\ref{fig:profiles(free)}) for the Ni-doped sample at 1.5 and 40~K.
Solid lines represent the result of fits by a Gaussian function.
The bottom figures show temperature dependence of intensity at the peak position.
Solid and dashed lines are guides to the eye and estimated backgrounds.
}
\label{fig:elastic}
\end{center}
\end{figure}

\subsection{\label{sec:results_3}Temperature dependence of spin excitations}

The temperature dependence of $\chi ''$($Q, \omega$) at $\omega =3$~meV is shown in Fig.~\ref{fig:kai(T-dep)_Zn} for the Zn-doped samples.
$\chi ''(Q, \omega)$ represents a fitted peak amplitude corrected with the thermal population factor.
$\chi ''(Q, \omega)$ of Zn:$y=0.004$ decreases below $T_{\rm c}$, suggesting the gap opening similar to the impurity-free sample.
In Zn:$y=0.008$ and Zn:$y=0.011$, the magnetic signal once starts decreasing around $T_{\rm c}$, which suggests the gap opening.
However, the signal increases again at lower temperature, indicating an emergence of the additional state.
We show a clear proof of the upturn behavior.
Figure~\ref{fig:profile_Zn11} presents peak profiles of Zn:$y=0.011$ at $T=10$, 15 and 20~K.
Apparently, the peak intensity at $T=15$~K is smaller than those at $T=10$ and 20~K.
This upturn behavior indicates two components of the spin excitations,
and we regard them as an additive state and the spin gap state.
We consider that the low-energy additive state
is not caused by the gap shift nor broadening and that it is a novel spin state induced by Zn.
Here we call the novel state an {\em in-gap} state, as defined by Kimura {\it et al.}~\cite{Kimura2003_a, Kimura2003_b}
Note that the temperature which the signals begin to decrease in Zn:$y=0.011$ is lower than that of Zn:$y=0.008$,
suggesting that the temperature of gap opening is associated with $T_{\rm c}$.
In Zn:$y=0.017$, $\chi ''$($Q, \omega$) shows no temperature dependence,
implying the reduction of spin gap energy and the development of the in-gap state.

In Fig.~\ref{fig:kai(T-dep)_Ni}, we show the temperature dependence of $\chi ''(Q, \omega)$ at $\omega =3$~meV for the Ni-doped samples.
In Ni:$y=0.009$, $\chi ''(Q, \omega)$ decreases below $T_{\rm c}$, suggesting a gap opening.
$\chi ''(Q, \omega)$ of Ni:$y=0.018$ shows no upturn as opposed to those of Zn:$y=0.008$ and Zn:$y=0.011$,
which indicates the absence of the in-gap state.
Thus, in the Ni-doped samples, the spin gap simply closes with increasing Ni,
which is not caused by the development of the in-gap state but by the gap shift or broadening.
%

\subsection{\label{sec:results_4}$Q$-$\omega $ dependence of incommensurate peaks}

In order to investigate the impurity effects on the magnetic peak profile in more detail,
we performed inelastic neutron scattering measurements under a high-resolution condition.
Figure~\ref{fig:profile(8meV)_high} shows peak profiles at $\omega =8$~meV for the impurity-free, the Zn-doped with Zn:$y=0.017$ and the Ni-doped with Ni:$y=0.029$ samples,
taken at 11~K, 4~K and 4~K, respectively.
In Zn:$y=0.017$, clear incommensurate magnetic peaks were observed,
and the peak width and position are almost identical to those of impurity-free sample.
In contrast, the magnetic peak width broadens in Ni:$y=0.029$ and the peak position
shifts toward the magnetic zone center (0.5, 0.5, 0).
In Fig.~\ref{fig:dispersion}, we plot $q$-$\omega $ dependence of peak position 
for the impurity-free, the Zn-doped (Zn:$y=0.011$, 0.017) and the Ni-doped (Ni:$y=0.018$, 0.029) samples.
Shaded regions represent FWHM of incommensurate magnetic peaks,
and a dashed line shows the peak positions for the impurity-free sample.
In the impurity-free sample, the peak positions and widths are independent of the excitation energy.
The peak positions and widths of the Zn-doped samples almost coincide with those of the impurity-free sample.
While in the Ni-doped samples, as the excitation energy increases, the peak positions apparently deviate from the dashed line and 
approach $q=0$.
Furthermore, the peak broadens more drastically with increasing energy.
These results suggest that the spin excitations of Ni-doped samples are dispersive and broaden with increasing energy.
These phenomena are characteristic of Ni and become marked with increasing the doping concentration.

\subsection{\label{sec:results_5}Elastic scattering}

We performed elastic scattering measurements for Zn:$y=0.017$ and Ni:$y=0.029$.
Figure~\ref{fig:elastic} shows the peak profiles at $T=1.5$ and 40~K
and the temperature dependence of the elastic peak intensity.
In Zn:$y=0.008$ and Zn:$y=0.011$, there is no observable signal, suggesting that the spin gap still opens.
However, a very sharp peak was clearly observed in Zn:$y=0.017$ at $T=1.5$~K and the signals appear below $T \sim 20$~K.
On the other hand, in the Ni-doped LSCO, we found a broader peak, which shows no temperature dependence.
Recently, Churei {\it et al.} have reported that no clear elastic peak was detected while a peak was observed at $\omega =0.4$~meV
for La$_{1.85}$Sr$_{0.15}$Cu$_{1-y}$Ni$_{y}$O$_{4}$ ($y=0.02$), 
with the instrumental energy resolution of $\sim 0.2$~meV. ~\cite{Churei2003}
In our case, the instrumental resolution is $\sim 1.2$~meV,
and we speculate that the observed ``elastic'' peaks arise from the inelastic signals which are detected due to a broad energy resolution.
We discuss these elastic signals from the viewpoint of spectral weight shift of spin excitation in Sec.~\ref{sec:discuss_1}.

\section{\label{sec:discuss}discussion}

\subsection{\label{sec:discuss_1}Differences between Zn and Ni doping effects}

We summarize impurity doping effects in Table~\ref{tab:comparison}.
In both Zn:$y=0.008$ and Zn:$y=0.011$, we observed an upturn behavior in the temperature dependence of $\chi ''(Q, \omega)$ at $\omega =3$~meV.
This result suggests the existence of two components; an in-gap state and a spin gap state.
On the other hand, there is no upturn for Ni-doping, indicating the absence of in-gap state.
%
The elastic signal induced by Zn is also different from those by Ni:
Zn induces a very sharp peak below $T \sim 20$~K and the signal exponentially increases with cooling,
while we found a broad peak which shows no temperature dependence in the Ni-doped LSCO.
Hirota {\it et al.} reported that Zn-doping of $y=0.012$ shifts the spectral weight of the spin excitations
from the inelastic to the quasi-elastic region at low temperature.~\cite{Hirota1998, Hirota2001}
We consider that the increase of elastic signal at low temperature is owing to the spectral-weight shift 
resulting from the development of the in-gap state.
As for Ni-doping, the absence of the in-gap state yields no shift of the spectral weight,
leading to no elastic signal.

\begin{table}
\caption{\label{tab:comparison}
Differences between Zn and Ni doping effects.
}
\begin{ruledtabular}
\begin{tabular}{c|c|c}
 & Zn & Ni \\ \hline
$\Delta T_{\rm c}$ & almost unchanged & broadening \\ \hline
- spin gap state - &  &  \\ 
$\bullet $ gap & slight gap shift & gap shift/broadening \\ 
$\bullet $ excitations & almost unchanged & enhance, dispersive \\ \hline
in-gap state & $\bigcirc $ & $\times $ \\ \hline
elastic signals & $\bigcirc $ & $\times $ \\ \hline
non-SC island & $\xi $ : large & $\xi $ : small or 0 \\ 
\end{tabular}
\end{ruledtabular}
\end{table}

As for the spin gap state in the Zn-doped LSCO, the gap slightly moves to the lower energy, but profiles of spin excitations are almost unchanged.
On the contrary, Ni changes the spin excitations drastically:
The spin gap shifts or broadens, and moreover, 
$\chi ''$ is enhanced and the spin excitations become dispersive.
Ni strongly affects not only the spin gap state but also the superconducting transition unlike Zn-doping.
The transition width $\Delta T_{\rm c}$ is almost constant as a function of the Zn concentration,
while $\Delta T_{\rm c}$ broadens with increasing Ni.
We speculate that the effects on the spin gap state are associated with the change of the electronic state of the superconductivity.

For impurity-doped systems, inhomogeneous pictures have been proposed~\cite{Nachumi1996, Adachi2004_a, Adachi2004_b};
impurities make the non-superconducting {\em island} with radius $\xi $ in the superconducting {\em sea}, and local moments reside in the island.
In our results, Zn-doping yields two components in magnetic excitations, 
and we regard them as the in-gap state in the island and the spin gap state in the sea.
We consider that the spin correlations in different islands are connected across the superconducting sea, which we call an {\em inter}-island correlation.
We discuss the inter-island correlation later.
In the Ni-doped LSCO, we observed no evidence for the in-gap state.
However, $\mu$SR studies and uniform susceptibility measurements suggest that 
Ni also induces local moments in a non-superconducting island similar to Zn.~\cite{Adachi2004_a, Adachi2004_b, Terascon1987, Xaio1990, Nakano1998}
Neutron scattering at a finite wave vector detects not local spin behaviors but spin correlations with a certain coherence.
If spin correlations are considerably short, we may not detect signals due to a too broadened peak profile.
In the case of Ni-doping, we speculate that spins only correlate within an individual island due to smaller $\xi$.
In fact, a recent $\mu$SR study of impurity-doped LSCO suggested that the island in a Ni-doped sample is smaller than that of a Zn-doped sample.~\cite{Adachi2004_b}
Thus, we conclude that the in-gap state is invisible and that we only observed the spin state of superconducting sea in the case of Ni-doping.

\begin{figure}
\begin{center}
\includegraphics[width=\hsize]{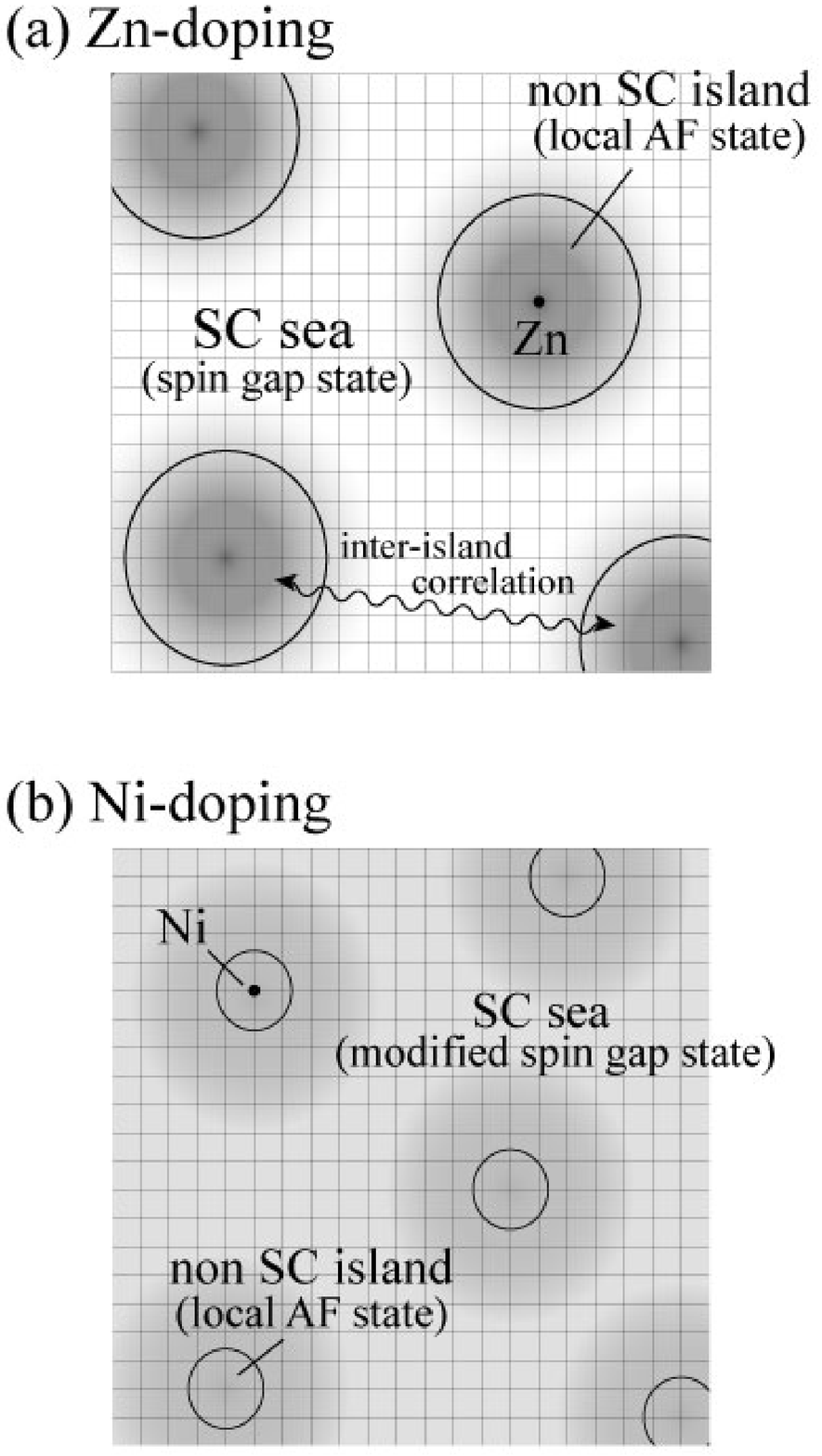}
\caption{
Schematic drawing of an inhomogeneous mixture of the superconducting (SC) sea with the spin gap state and the non-superconducting islands with the local AF correlation.
(a) and (b) correspond to the cases of Zn-doping and Ni-doping.
Refer to the text for the meaning of inter-island correlation and modified spin gap state.
}
\label{fig:swiss_cheese}
\end{center}
\end{figure}

We schematically show our concept in Fig.~\ref{fig:swiss_cheese}, from a viewpoint of the inhomogeneous picture.
Zn induces non-superconducting islands and hardly affects the superconducting sea, so that the spin gap state and $\Delta T_{\rm c}$ are almost unchanged.
While Ni primarily effects on the sea.
In other word, Ni affects the electronic and magnetic state of the superconductivity,
leading to the modified spin gap state and the broadening of $\Delta T_{\rm c}$.
We believe that 
the gap is {\em filled} by the development of the in-gap state and the slight gap shift in the case of Zn-doping,
while the gap shift or broadening simply {\em closes} the gap for Ni-doping.

\subsection{Zn doping effects; the in-gap state}

A primary effect by Zn on the spin excitations is the appearance of in-gap state.
Most obvious evidence for the in-gap state is the upturn in the temperature dependence of $\chi ''(Q, \omega)$ at $\omega =3$~meV.
Previously Kimura {\it et al.} reported the upturn behavior in Zn:$y=0.008$.~\cite{Kimura2003_a, Kimura2003_b}
In the present study, we also confirmed the upturn behavior in Zn:$y=0.011$.
%
%
In YBa$_{2}$(Cu$_{1-y}$Zn$_y$)$_{3}$O$_{6.97}$ ($y=0.02$), similar results have been reported by Sidis {\it et al.}~\cite{Sidis1996}
Their inelastic neutron scattering measurements showed that magnetic signals emerge around $\omega \sim 9$~meV while the spin gap is retained,
and that $\chi ''$ at $\omega =10$~meV is almost temperature independent below $T_{\rm c}$.
Their results indicate the coexistence of the in-gap state with the spin gap state at low temperature,
signifying that the appearance of in-gap state is a universal phenomenon in Zn-doped high-$T_{\rm c}$ cuprates.
Note that the in-gap state appears at the same $Q$ position as that of the normal and high energy states for both LSCO and YBCO.
Therefore, Zn locally slows spin fluctuations without modifying the AF wave vector.
In other words, Zn seems to act as a pinning center of AF spin fluctuations.

As for Zn:$y=0.008$, the in-plane spin correlation length $\xi_{ab}$($\geq $~80~\AA) of the in-gap state is much longer than $R_{\rm Zn-Zn}$($\sim$~42~\AA),
which corresponds to the mean distance between the nearest-neighbor Zn ions.
These results indicate that the spin correlations of the in-gap state do not result from the local AF coherence in an individual island,
but from the inter-island correlation.
One may argue that it is caused by an overlap of islands.
However, our results show the coexistence of the spin gap and the in-gap state (See Fig.~\ref{fig:kai(T-dep)_Zn}),
suggesting a phase separation of islands and sea.
Therefore we conclude that induced moments in an island correlate with those in spatially separated islands.

\subsection{A threshold in Ni doping effects}

We observed that the amplitude of $\chi ''(\omega )$ for Ni:$y=0.018$ and Ni:$y=0.029$ is much larger than those for $y=0$ and Ni:$y=0.009$,
signifying that there exists a certain threshold between Ni:$y=0.009$ and Ni:$y=0.018$.
We speculate that the threshold corresponds to a critical Ni concentration $y_0$ where the valence of Ni ions changes;
%
Nakano {\it et al.} have reported that no Curie term appears in the uniform susceptibility measurements 
below the critical Ni concentration $y_0$~\cite{Nakano1998}.
They also reported that the characteristic temperature which $\chi $ exhibits a broad peak, so-called $T_{\rm max}$, increases with increasing Ni below $y_0$.
On the other hand, the Curie term appears and $T_{\rm max}$ does not change above $y_0$.
If the valence of Ni ions are not divalent, the effective hole concentration $p$ should change,
leading to the variation of $T_{\rm max}$ which is susceptible to $p$.
From these results, they conclude that Ni ions are substituted for Cu ions
as Ni$^{3+}$ ($S=1/2$ : low-spin state) below $y_0$ and Ni$^{2+}$ ($S=1$) above $y_0$;
Ni$^{3+}$ has the spin of $S=1/2$ as well as that of Cu$^{2+}$, indicating that the Ni$^{3+}$ substitution hardly affects the spin correlation in the CuO$_2$ plane.
In our result, $\chi ''(\omega )$ of Ni:$y=0.009$ is almost same as those of $y=0$ except for a slight gap broadening.
We consider that this behavior of Ni:$y=0.009$ is consistent with the absence of the Curie term and the reduction of $p$ below $y_0$ ($\sim 0.01$ for $x=0.15$).

\subsection{Spin excitations: a possible scaling with $T_{\rm c}$}

\begin{figure}[t]
\begin{center}
\includegraphics[width=\hsize]{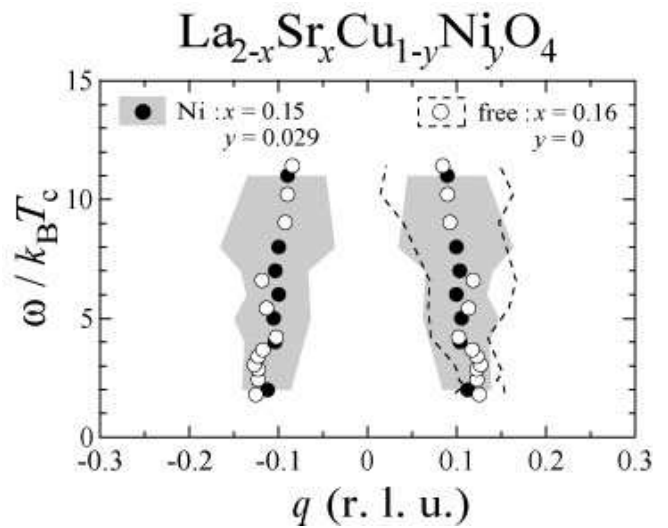}
\caption{
Dispersions of spin excitations scaled by $k_{\rm B}T_{\rm c}$ for La$_{1.85}$Sr$_{0.15}$Cu$_{1-y}$Ni$_{y}$O$_{4}$ (Ni:$y=0.029$) at 4~K
and La$_{1.84}$Sr$_{0.16}$CuO$_{4}$ (free:$x=0.16$) at 10~K~\cite{Christensen2004}.
$q$ denotes the propagation vector of spin correlation.
Closed (open) circles and shaded (dashed) regions represent the peak position and FWHM for Ni:$y=0.029$ (free:$x=0.16$), respectively.
}
\label{fig:dispersion_scale}
\end{center}
\end{figure}

%
Recently, quite similar spin excitations are discovered in both YBCO~\cite{Hayden2004} and La$_{2-x}$Ba$_{x}$CuO$_{4}$ (LBCO)~\cite{Tranquada2004_b}
by a time-of-flight neutron spectrometer which covers a wide range of energy:
The low-energy spin excitations form incommensurate peaks, and the peaks disperse inwards toward $Q_{\rm AF}$.
At a characteristic energy, which is called $E_r$ in YBCO, the peak is found at a commensurate position $Q=Q_{\rm AF}$.
In higher energy region, the excitations disperse outwards and form a square-shaped continuum.
Furthermore, the dispersive excitations at low-energy are observed in LSCO ($x=0.16$),~\cite{Christensen2004}
though the commensurate peak and the dispersion in higher energy region are not yet confirmed.
These results suggest that such dispersive excitations are a common feature in the high-$T_{\rm c}$ cuprates.
In our study, we measured spin excitations in the low energy region $\omega \leq 11$~meV, so that we cannot detect the dispersive excitations 
in the impurity-free and the Zn-doped samples.
However, to our surprise, the present results show that the spin excitations of Ni-doped samples are dispersive and broaden with increasing energy.
These behaviors are very similar to those of impurity-free LSCO~\cite{Christensen2004}, when the energy scale of spin excitations is adjusted.
We plot the dispersion of spin excitations scaled to $T_{\rm c}$ as shown in Fig.~\ref{fig:dispersion_scale}.
This behavior indicates that Ni reduces the energy scale of spin excitations which associates with $T_{\rm c}$.
A previous neutron scattering study of YBCO reported that 
Ni reduces $E_r$ with conserving the ratio $E_r/T_{\rm c}$, while such a shift is much smaller in the Zn-doped sample.~\cite{Sidis2000}
Furthermore, Cu NQR study of Ni-doped YBCO has shown that all the 1/$T_1$ data in the normal state are on a universal curve,
when 1/$T_1$ is plotted against $t=T/T_{\rm c}$.
Since 1/$T_1$ is related to $\chi'' (Q, \omega )$,
the universality of 1/$T_1$ implies that AF spin fluctuations are scaled to $T_{\rm c}$ upon Ni-doping.~\cite{Tokunaga1997}
These results are consistent with ours, and we consider that these behaviors are consequences of the reduction of energy scale of spin excitations by Ni.
However, we cannot affirm how precisely the spin excitations are scaled to $T_{\rm c}$, 
because we measured in the narrow energy region ($\omega \leq 11$~meV).
Further inelastic neutron scattering studies in higher energy region are strongly required.



\section{\label{sec:conclusion}conclusion}

Zn-doping induces local moments within individual non-superconducting islands and the islands are spatially separated from the superconducting sea, 
which is hardly affected by Zn.
At low temperature, the local moments among different islands correlate, leading to the emergence of an in-gap state.
The in-gap state appears at the same $Q$ position in the normal and high energy states, and Zn seems to act as a pinning center of AF spin fluctuations.
We consider that Zn reduces the volume fraction of the superconducting regions, which results in the reduction of $T_{\rm c}$.

We observed no evidence for the in-gap state in the case of Ni-doping.
We consider that the invisibility of the in-gap state arises from undeveloped spin correlations among different islands,
because the islands are smaller than those of Zn-doping.
Instead, Ni primarily effects on the superconducting sea;
the spin gap shifts or broadens, and Ni makes the spin excitations get broader and dispersive.
These behaviors are recognized as a consequence of the reduction of energy scale of spin excitations.
We believe that the reduction of energy scale is relevant to the suppression of $T_{\rm c}$.

\begin{acknowledgments}
We are grateful to M. Matsuura for experimental supports and critical discussions throughout the project.
We are indebt to S. Hosoya and M. Onodera for technical helps in the crystal growth.
Neutron scattering experiments were supported by T. Asami, Y. Kawamura, K. Nemoto, K. Ohoyama, and S. Watanabe.
We also thank T. Adachi, T. Churei, Y. Endoh, M. Fujita, H. Hiraka, K. Ishida, Y. Itoh, K. Iwasa, C. H. Lee, M. Matsuda, Y. Murakami, Y. Noda, and K. Yamada
for stimulating discussions.
This work was supported in part by a Grant-In-Aid for Scientific Research (B) (15340109, 2003-2004)
and a Grant-In-Aid for Encouragement of Young Scientist (13740198, 2001-2002)
from the Japanese Ministry of Education, Culture, Sports, Science and Technology.

\end{acknowledgments}

\bibliography{PRB}

\end{document}